%% file: seqEpaper.tex
\documentclass[useAMS,usenatbib]{mn2e}
\usepackage[dvips]{graphicx}
\begin{document}

\title[Ellipsoidal Variability and the Difference between Sequence D and E Red Giants]{Ellipsoidal Variability and the Difference between Sequence D and E Red Giants}
\author[C. P. Nicholls et al.]{C. P. Nicholls$^{1}$\thanks{E-mail:
nicholls@mso.anu.edu.au (CPN); wood@mso.anu.edu.au (PRW);  m.cioni@herts.ac.uk
(M-RLC)}, P. R. Wood$^{1}$\footnotemark[1] and M.-R. L. Cioni$^{2}$\footnotemark[1]\\
$^{1}$Research School of Astronomy and Astrophysics, Australian National
University, Cotter Road, Weston Creek ACT 2611, Australia\\
$^{2}$Centre for Astrophysics Research, University of Hertfordshire, College Lane, Hatfield, AL10 9AB, UK}

\date{Accepted 2010 February 18. Received 2010 February 18; in original form 2009 December 10.}

\pagerange{\pageref{firstpage}--\pageref{lastpage}} \pubyear{2010}

\maketitle

\label{firstpage}

\begin{abstract}

We have studied a sample of Large Magellanic Cloud red giant binaries that lie on sequence E in the period--luminosity plane. We show that their combined light and velocity curves unambiguously demonstrate that they are binaries showing ellipsoidal variability.
By comparing the phased light and velocity curves of both sequence D and E variables, we show that the sequence D variation -- the Long Secondary Period -- is not caused by ellipsoidal variability. We also demonstrate several further differences between stars on sequences D and E. These include differences in velocity amplitude, in the distribution of eccentricity, and in the correlations of velocity amplitude with luminosity and period. We also show that the sequence E stars, unlike stars on sequence D, do not show any evidence of a mid-infrared excess that would indicate circumstellar dust.

\end{abstract}


\section{Introduction}

Long Period Variables (LPVs) are known to fall on different sequences in the period--luminosity plane \nocite{wood99mn} \nocite{ogle04} \citep[Wood et al.\ 1999; Soszy\'nski et al.\ 2004a;][]{ita04, fraser05, oglep-l, fraser08}. One of these sequences, known as sequence E, is thought to consist of red giants in close binary systems showing ellipsoidal variability, although this has not been unambiguously demonstrated. A small number of the binaries appear to be eclipsing and others appear to have unexpectedly eccentric orbits, based on their light curve shape \nocite{ogleellipsoidal} (Soszy\'nski et al. 2004b).

A star in a close binary will have its Roche Lobe distorted by the tidal influence of its orbiting companion. When the star begins to fill its Roche Lobe it takes on an elongated, or ellipsoidal shape, becoming an ellipsoidal variable. The velocity variations of such stars are dominated by the orbital motion, but light variability is caused mainly by the change in apparent surface area as the star orbits around its companion. Because of this, the light curve of an ellipsoidal variable shows two maxima and minima per orbit -- two cycles for every one cycle of the velocity curve. This is an easy way to unambiguously identify ellipsoidal variables and we use this test here.

Four of the other sequences of LPVs -- A, B, C$'$ and C -- are known to harbour radially pulsating variables \nocite{wood99mn} (Wood et al.\ 1999). A fifth sequence, sequence D, contains stars which show Long Secondary Periods. The origin of Long Secondary Periods (LSPs) remains something of a mystery, though several attempts have been made to discover it \nocite{wood99mn} \citep[Wood et al.\ 1999;][]{hinkle02,olivierwood03,sequenceDstars,seqDpaper}. Due to their overlap in the period--luminosity diagram, some authors \nocite{ogleellipsoidal} \citep[e.g. Soszy\'nski et al. 2004b,][]{soszynski07} have suggested that stars on sequences D and E may be fundamentally the same -- binaries showing ellipsoidal variability.

Here we study a sample of sequence E binaries. In particular we present new radial velocity data derived from VLT spectra, which we use alongside MACHO light curves. We show that the variations of stars on sequences D and E are caused by different mechanisms. Preliminary results for three stars are given in \cite{betsy}.

\section{Observations and Data Reduction}

The observations and data reduction for this sample are the same as for our recently published sequence D sample \citep{seqDpaper}. The spectra were taken using the FLAMES/ GIRAFFE spectrograph (Pasquini et al.\ 2002) \nocite{pasquinimn} on the European Southern Observatory's Very Large Telescope (VLT), on 21 nights from 2003 November to 2006 March. Radial velocities were calculated via cross-correlation with the \textsc{iraf} task \emph{fxcor}. The reader is referred to \cite{seqDpaper} for details. Table~\ref{vtable} shows the radial velocities calculated for part of our sample for a few dates. The full table, with radial velocities of our whole sample for all dates, is available online.

\begin{table}
\centering
\caption{Radial Velocities of Sequence E Stars. Stars are identified by their MACHO numbers.}
\label{vtable}
\begin{tabular}{lrrrr}
\hline
\multicolumn{1}{c}{HJD}  &  \multicolumn{1}{c}{77.7429.189}  &  \multicolumn{1}{c}{77.7548.68}  &  \multicolumn{1}{c}{77.7672.98}  &  \multicolumn{1}{c}{77.7673.79} \\
\hline                                                                           
2954.8508  &   0.00  &       247.1836  &   270.6885  &   265.7394  \\
3005.8604  &  0.00  &       242.5239  &   260.0903  &   258.5819   \\
3067.6028  &  0.00  &       237.8895  &   225.7872  &   239.5893   \\
3091.5571  &  0.00  &       238.4238  &   251.1337  &   265.9271   \\
3280.8611  &  272.6101  &   255.7782  &   284.0382  &   244.0299 \\
\hline
 \end{tabular}
 \end{table}

\section{Results}
\label{results}

\begin{figure*}
\begin{center}
\includegraphics[width=0.9\textwidth]{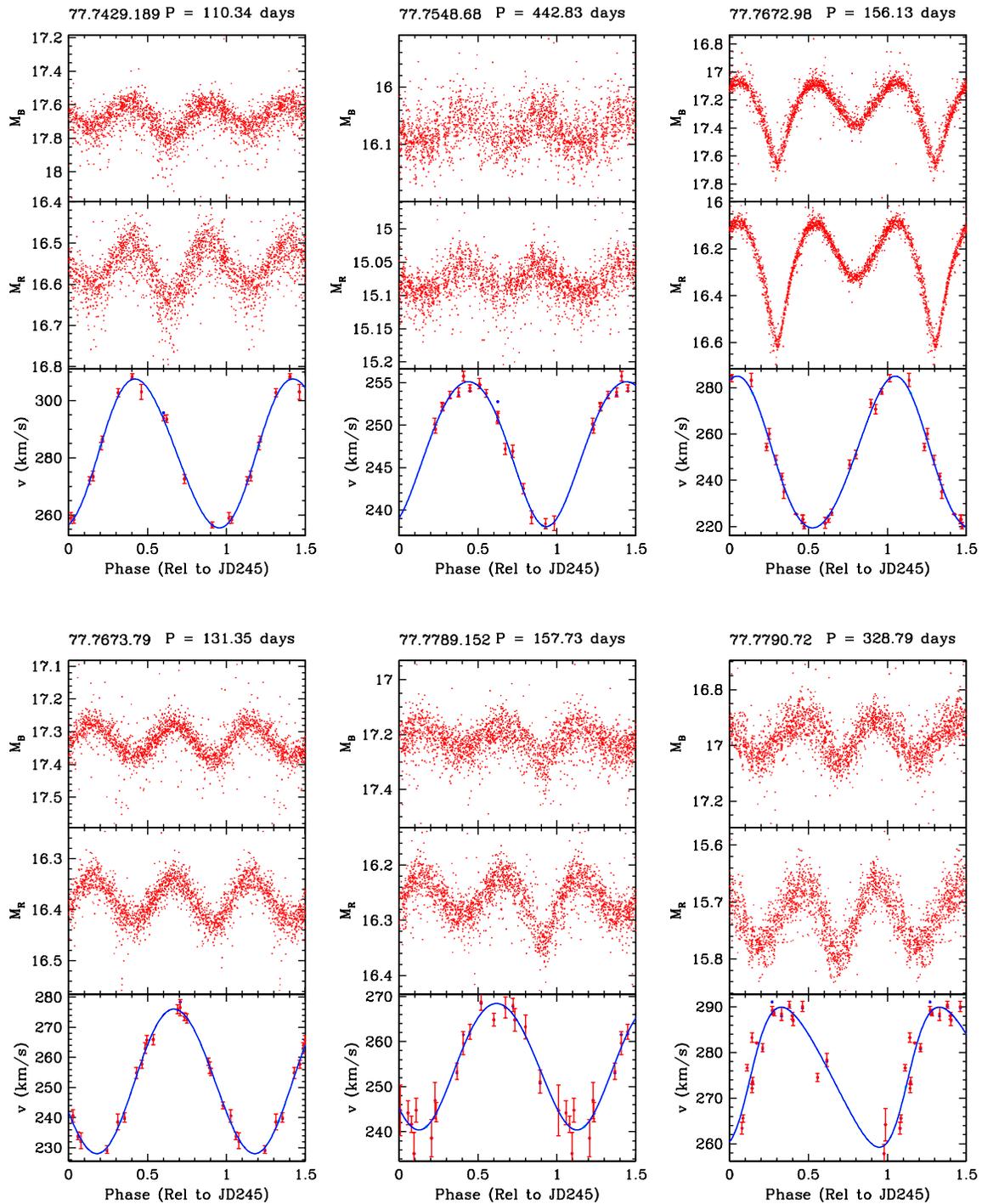}
\end{center}
\caption[Phased light and velocity.]{Phased MACHO blue ($M_B$) and red ($M_R$) light curves and phased radial velocity curves for our sequence E stars. The blue line shows the binary fit to the velocity curve.}
\label{phased1}
\end{figure*}

\begin{figure*}
\begin{center}
\includegraphics[width=0.9\textwidth]{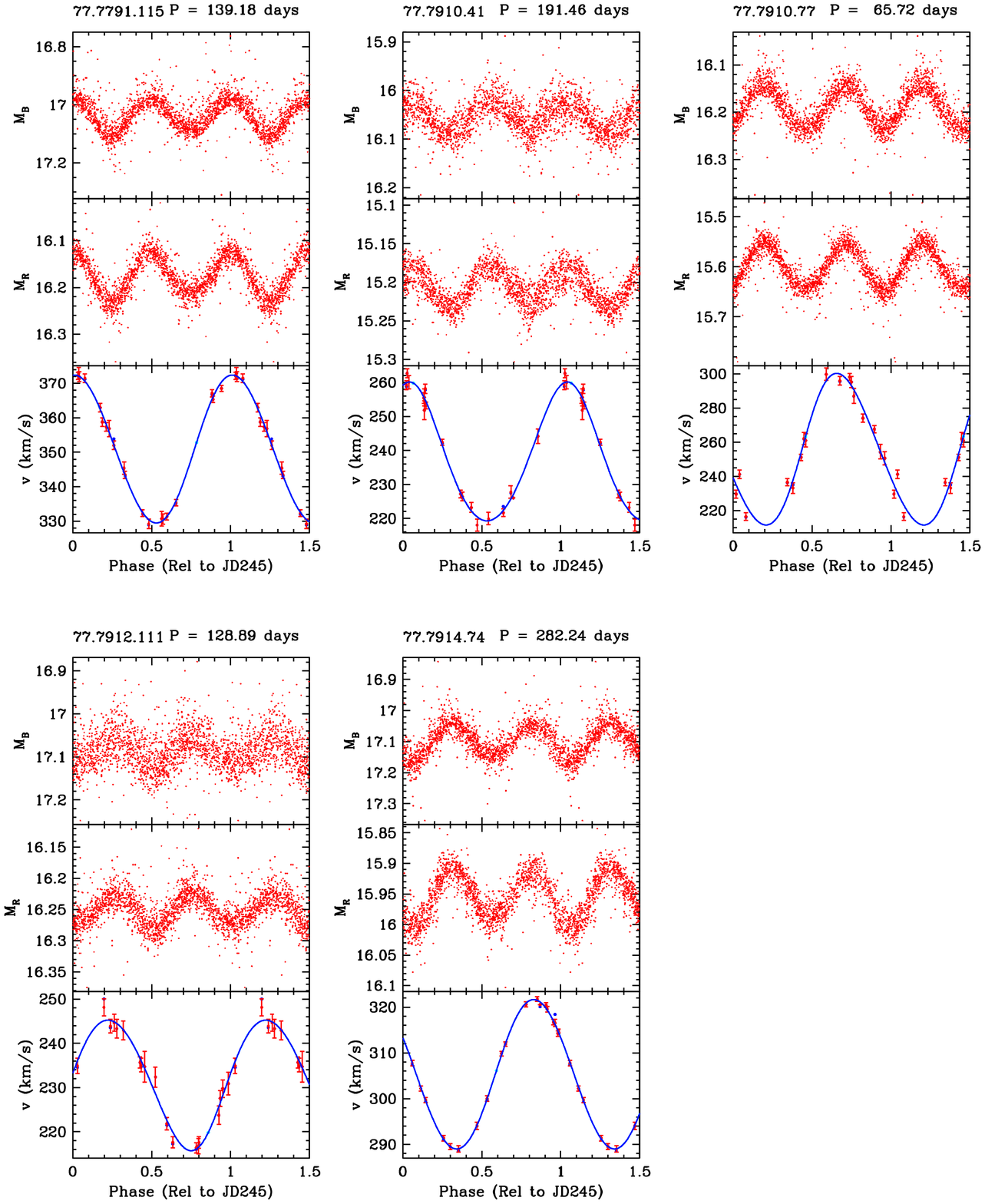}
\end{center}
\caption[Phased light and velocity.]{Same as Fig.~\ref{phased1}.}
\label{phased2}
\end{figure*}

Plots of phased light and velocity variations for all the stars in our sequence E sample are shown in Figs.~\ref{phased1} and~\ref{phased2}. A clear doubling of the phased light curve with respect to the phased velocity curve can be seen for all stars in our sample. This behaviour unambiguously demonstrates that these stars are ellipsoidal variables. Many of the stars show lightcurves with equal maxima and unequal minima, which suggests that one end of the ellipsoid is hotter and brighter than the other. From Figs.~\ref{phased1} and~\ref{phased2} we see that the deepest minimum occurs at mean velocity during decreasing radial velocity, i.e. when the red giant is behind its companion. Therefore the inner end of the red giant ellipsoid -- the end closest to the companion -- is the cooler end, and the side furthest from the companion is hotter. This is most likely due to gravity darkening towards the companion.

\begin{figure}
\begin{center}
\includegraphics[width=0.5\textwidth]{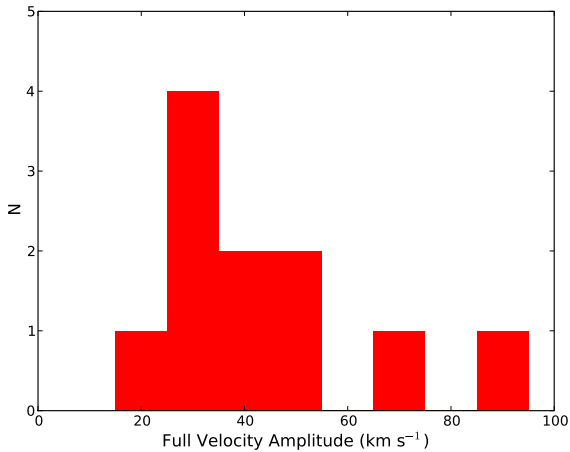}
\end{center}
\caption[Velocity amplitude histogram.]{A histogram of the velocity amplitude for our sample.}
\label{vhist}
\end{figure}

A histogram of the velocity amplitudes of our sample is shown in Fig.~\ref{vhist}. The majority of values lie between 15 and $55\ \rm{km\,s^{-1}}$, and the mean velocity amplitude is $43.3\ \rm{km\,s^{-1}}$. 

We have made a binary fit to the velocity curves of our sequence E red giants \citep[see section 2.5 of ][]{seqDpaper}. We calculated the mass function, 
\begin{equation}
f(m) = \frac{K^3 P}{2 \pi G} = \frac{m^3 \sin ^3 i}{(m+M)^2} ,
\end{equation} 
using the observed values for period ($P$) and velocity semiamplitude ($K$) (Table~\ref{orbital}). Using the calculated mass function of our binary fit and assuming a total system mass of $2\ \rm{M_{\odot}}$, we calculated an estimated companion mass for each of the stars in our sample. As expected, for most systems the companion is less massive than the red giant, but for two stars this resulted in companions with significantly higher mass than the red giant. This is unlikely since the more massive companion should evolve to the red giant stage first, and even if the current red giant was originally the less massive star, its companion would most likely now be a white dwarf of lower mass than the red giant. In these two cases we chose the total mass to be such that the red giant and its companion had equal mass. This results in higher masses for these stars (total system masses of $\sim 4.5 \rm{M_{\odot}}$.) The mass estimates are shown in Table~\ref{orbital}, alongside the other parameters of the binary fit to the velocity. Here $\gamma$ is the system velocity, $e$ is the eccentricity, $\omega$ is the angle of periastron, $T$ is the date of periastron, $a \sin i$ is the semimajor axis of the red giant's orbit, $f(m)$ is the mass function, $M$ is the mass of the red giant, and $m$ the mass of its companion. We assume $\sin i = 1$ when calculating the masses. The errors for each element are shown on the line below each star.

\begin{table*}
\centering
\begin{minipage}{160mm}
\caption{Orbital Elements for Sequence E Stars}
\label{orbital}
\begin{tabular}{lrrrrrrrrrr}
\hline
\multicolumn{1}{c}{Star}  &  \multicolumn{1}{c}{$\gamma$}  &  \multicolumn{1}{c}{$K$}  &  \multicolumn{1}{c}{$e$}  &  \multicolumn{1}{c}{$\omega$}  &  \multicolumn{1}{c}{$T$}  &  \multicolumn{1}{c}{$P$}  &  \multicolumn{1}{c}{$a \sin i$}  &  \multicolumn{1}{c}{$f(m)$}  &  \multicolumn{1}{c}{$M$%
\footnote{Here we assume $\sin i = 1$.}}  &  \multicolumn{1}{c}{$m$}\\
  &  \multicolumn{1}{c}{($\rm{km\,s^{-1}}$)}  &  \multicolumn{1}{c}{($\rm{km\,s^{-1}}$)}  &  &  \multicolumn{1}{c}{(deg)}  &  \multicolumn{1}{c}{(HJD)}  &  \multicolumn{1}{c}{(days)}  &  \multicolumn{1}{c}{($\rm{R_{\odot}}$)}  &  \multicolumn{1}{c}{($\rm{M_{\odot}}$)}  &  \multicolumn{1}{c}{($\rm{M_{\odot}}$)}  &  \multicolumn{1}{c}{($\rm{M_{\odot}}$)}\\
\hline
77.7429.189 &  281.40 &  25.99 & 0.05 & 275.38 &  3332.5 & 110.34 &  56.58 &   0.2003 &  1.07 &  0.93 \\
 & $\pm$  0.57 & $\pm$  0.80 & $\pm$ 0.03 & $\pm$ 28.52 & $\pm$   8.8 &  & $\pm$ 1.75 & $\pm$ 0.0186 &  &  \\
77.7548.68 &  247.20 &   8.54 & 0.07 & 170.67 &  3502.6 & 442.83 &  74.55 &   0.0285 &  1.52 &  0.48 \\
 & $\pm$  0.19 & $\pm$  0.24 & $\pm$ 0.03 & $\pm$ 25.63 & $\pm$  32.9 &  & $\pm$ 2.22 & $\pm$ 0.0024 &  &  \\
77.7672.98 &  250.71 &  32.77 & 0.06 & 38.99 &  2989.4 & 156.13 & 100.92 &   0.5679 &  2.27 &  2.27 \\
 & $\pm$  0.75 & $\pm$  1.10 & $\pm$ 0.03 & $\pm$ 29.11 & $\pm$  12.3 &  & $\pm$ 3.42 & $\pm$ 0.0575 &  &  \\
77.7673.79 &  252.37 &  24.01 & 0.02 & 236.49 &  3064.4 & 131.35 &  62.30 &   0.1887 &  1.09 &  0.91 \\
 & $\pm$  0.41 & $\pm$  0.59 & $\pm$ 0.02 & $\pm$ 61.77 & $\pm$  22.6 &  & $\pm$ 1.55 & $\pm$ 0.0140 &  &  \\
77.7789.152 &  255.04 &  14.03 & 0.05 & 198.55 &  3024.2 & 157.73 &  43.67 &   0.0451 &  1.44 &  0.56 \\
 & $\pm$  0.74 & $\pm$  0.97 & $\pm$ 0.08 & $\pm$ 90.81 & $\pm$  40.5 &  & $\pm$ 3.05 & $\pm$ 0.0094 &  &  \\
77.7790.72 &  275.00 &  15.35 & 0.19 & 262.20 &  3326.7 & 328.79 &  97.85 &   0.1167 &  1.22 &  0.78 \\
 & $\pm$  0.83 & $\pm$  0.92 & $\pm$ 0.06 & $\pm$ 13.11 & $\pm$  10.2 &  & $\pm$ 6.02 & $\pm$ 0.0214 &  &  \\
77.7791.115 &  350.88 &  21.47 & 0.03 & 275.04 &  3032.0 & 139.18 &  59.02 &   0.1429 &  1.17 &  0.83 \\
 & $\pm$  0.30 & $\pm$  0.35 & $\pm$ 0.02 & $\pm$ 40.51 & $\pm$  16.3 &  & $\pm$ 0.97 & $\pm$ 0.0070 &  &  \\
77.7910.41 &  238.06 &  20.41 & 0.08 & 10.03 &  3075.8 & 191.46 &  76.94 &   0.1673 &  1.13 &  0.87 \\
 & $\pm$  0.55 & $\pm$  0.66 & $\pm$ 0.04 & $\pm$ 21.82 & $\pm$  11.8 &  & $\pm$ 2.53 & $\pm$ 0.0163 &  &  \\
77.7910.77 &  255.34 &  44.35 & 0.08 & 279.65 &  2987.6 &  65.72 &  57.38 &   0.5890 &  2.36 &  2.36 \\
 & $\pm$  1.67 & $\pm$  2.83 & $\pm$ 0.04 & $\pm$ 38.55 & $\pm$   7.1 &  & $\pm$ 3.67 & $\pm$ 0.1131 &  &  \\
77.7912.111 &  231.16 &  14.76 & 0.06 & 223.73 &  3075.1 & 128.89 &  37.50 &   0.0428 &  1.44 &  0.56 \\
 & $\pm$  0.45 & $\pm$  0.62 & $\pm$ 0.04 & $\pm$ 41.73 & $\pm$  14.7 &  & $\pm$ 1.57 & $\pm$ 0.0054 &  &  \\
77.7914.74 &  305.31 &  16.33 & 0.02 & 272.80 &  3271.8 & 282.24 &  91.04 &   0.1276 &  1.20 &  0.80 \\
 & $\pm$  0.11 & $\pm$  0.14 & $\pm$ 0.01 & $\pm$ 24.26 & $\pm$  18.9 &  & $\pm$ 0.79 & $\pm$ 0.0033 &  &  \\
 \hline
 \end{tabular}
 \end{minipage}
 \end{table*}

The eccentricity of a binary system is expected to decrease over time due to circularising tidal forces \citep{zahn}, provided no mechanism is in place that will stabilise or increase the orbital eccentricity. Thus binaries with one or more red giant components are expected to describe relatively circular orbits. We confirmed this theory by calculating the circularisation time for our stars, using the formula given in \cite{soker00}. The median circularisation time of our sample is only $\sim$ 3500 y. For comparison, the time a $1\ \rm{M_{\odot}}$, $Z=0.008$ red giant takes to double its radius when $R\sim 30\ \rm{R_{\odot}}$ is $\sim 2 \times 10^7$ y \citep[using the evolutionary tracks of][]{girardi}. The value of $30\ \rm{R_{\odot}}$ was chosen as a typical Roche Lobe radius for these stars (see the velocity amplitude calculations given later in this section). A doubling of the radius is an estimate of the evolution time over which the tides have had time to act. Ten of the eleven stars in our sample have $e < 0.1$, and the mean eccentricity of the sample is 0.07. Therefore the majority of our sample of red giant binaries have fairly circular orbits, as expected. However some sequence E stars do show confoundingly high eccentricities \nocite{ogleellipsoidal} (see Soszy\'nski et al. 2004b), and this is a problem that must be solved in the future.

\begin{figure}
\begin{center}
\includegraphics[width=0.5\textwidth]{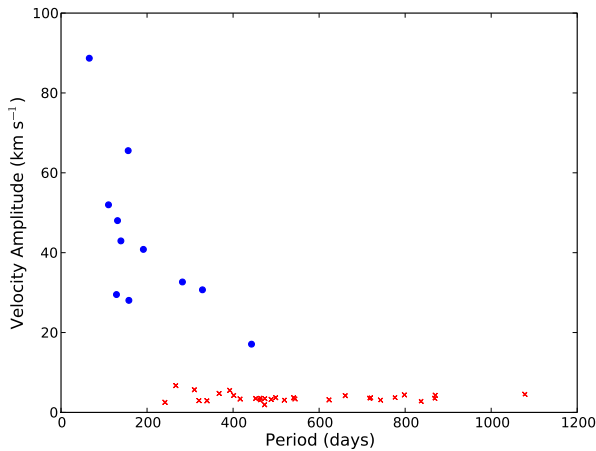}
\end{center}
\caption[Velocity amplitude plotted against orbital period.]{Full velocity amplitude plotted against orbital period for our sequence E sample (blue points) and the sequence D sample of \cite{seqDpaper} (red crosses).}
\label{vamp-p}
\end{figure}

The relation between velocity amplitude and period for our sample and for the sequence D sample of \cite{seqDpaper} is plotted in Fig.~\ref{vamp-p}. In a binary system, for given masses, a longer period means a wider orbit, and thus a smaller amplitude in velocity variation. Fig.~\ref{vamp-p} shows velocity amplitude decreases with increasing period for the sequence E stars, as expected. However the sequence D stars show the same velocity amplitudes for all periods and do not follow the expected binary relation. A vivid demonstration of the disparity between these two samples is found in the region where their periods overlap. At periods between 200 and 450 days, the difference in velocity amplitude between the sequence D and sequence E samples is as much as 30 $\rm{km\,s^{-1}}$.

\begin{figure}
\begin{center}
\includegraphics[width=0.5\textwidth]{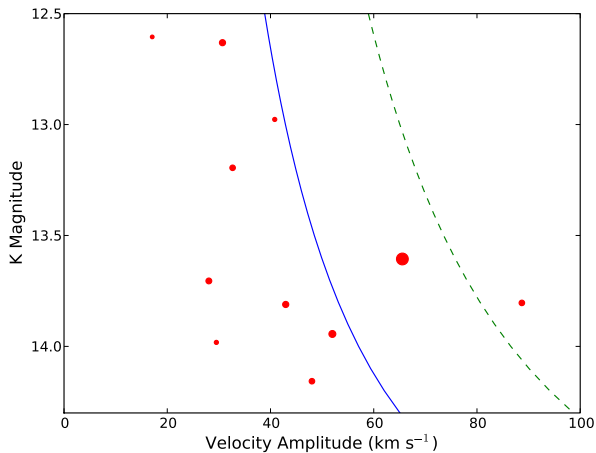}
\end{center}
\caption[$K$ magnitude plotted against velocity amplitude.]{$K$ magnitude (from 2MASS) plotted against velocity amplitude for our sequence E stars. The point size shows relative light amplitude in MACHO red. The solid blue curve gives the upper velocity limit for equal-mass components of $1\ \rm{M_{\odot}}$, and the dashed green curve gives the upper limit for equal-mass components of $2.3\ \rm{M_{\odot}}$. See text for details.}
\label{k-vamp}
\end{figure}

Ellipsoidal variability is only visible once a star has substantially filled its Roche Lobe. Stars in wider orbits (with lower velocity amplitudes) will fill their Roche Lobes when they are further up the giant branch, and thus will be more luminous than stars in closer orbits. For a given luminosity (and radius) there is a range of velocity amplitudes the star may have, the maximum of which is dictated by the size of the closest possible orbit the companion can occupy. Therefore, for our ellipsoidal variables we expect that the maximum velocity amplitude should decrease with increasing luminosity, but that stars may occupy velocity amplitudes below the maximum for a given luminosity. This is shown in Fig.~\ref{k-vamp} which gives a plot of $K$ magnitude against velocity amplitude for our sequence E stars. In order to define the upper limit of velocity amplitude for a given luminosity, we calculated the minimum orbital separation for a theoretical sample of binaries with equal mass components (mass ratios of $q = 1$) in which the red giant is filling its Roche Lobe. Using the approximation given in \cite{eggleton}, we calculated the Roche Lobe radius $r_{L}$ to be $0.37$, in units of orbital separation. Therefore the minimum orbital separation for a Roche-Lobe filling binary with $q = 1$ is $a=\frac{r_{L}}{0.37}=2.7r_{L}$. We substitute for $r_{L}$ the radius expected for a given luminosity in these stars, calculated from a fit made to the radius--$K$ magnitude data for our sample. Finally we calculated the maximum velocity amplitude assuming a circular orbit and components of mass $1\ \rm{M_{\odot}}$. This velocity amplitude upper limit is shown by the solid blue line in Fig.~\ref{k-vamp}.

Most stars lie where expected in Fig. ~\ref{k-vamp} but two stars lie above the upper limit line. These are the two systems with higher mass components, mentioned earlier. As both these systems have components of around $2.3\ \rm{M_{\odot}}$, we also calculated the maximum velocity amplitude for stars of this mass, and this is shown by the green dashed line in Fig.~\ref{k-vamp}. One star lies above this limit. It appears somewhat atypical, as it has a significantly higher effective temperature than the rest of the sample ($\sim5200 K$ compared to the median $4200 K$) and hence a smaller radius. It therefore does not overflow its Roche Lobe as Fig.~\ref{k-vamp} suggests. 

Table~\ref{properties} gives the radius, luminosity, effective temperature and orbital separation data for our sample. For the majority of stars, these properties were calculated using the 2MASS $J$ and $K$ magnitudes. However for the star 77.7910.77, which lies outside the green dashed limit in Fig.~\ref{k-vamp}, the properties were calculated from OGLE $V-I$ values, which are more reliable for warmer stars. Due to this star's warmer temperature, we suspect it could be a helium core burning clump star or a blue loop star.

\begin{table}
\centering
\caption{Sequence E Star Properties}
\label{properties}
\begin{tabular}{lrrrr}
\hline
\multicolumn{1}{c}{Star}  &  \multicolumn{1}{c}{$R$}  &  \multicolumn{1}{c}{$L$}  &  \multicolumn{1}{c}{$T_{\rm{eff}}$}  &  \multicolumn{1}{c}{$a$}\\
  &  \multicolumn{1}{c}{($\rm{R_{\odot}}$)}  &  \multicolumn{1}{c}{($\rm{L_{\odot}}$)}  & \multicolumn{1}{c}{(K)}  &  \multicolumn{1}{c}{($\rm{R_{\odot}}$)}\\
\hline
77.7429.189 &  48.92 &  593.16 & 4071.43 & 121.83 \\
77.7548.68 &  92.03 & 1917.72 & 3980.39 & 307.67 \\
77.7672.98 &  56.13 &  869.34 & 4181.83 & 201.84 \\
77.7673.79 &  42.24 &  591.59 & 4378.57 & 136.84 \\
77.7789.152 &  52.85 &  841.10 & 4274.26 & 154.60 \\
77.7790.72 &  92.71 & 1736.69 & 3868.59 & 252.27 \\
77.7791.115 &  49.66 &  805.48 & 4362.15 & 142.23 \\
77.7910.41 &  72.37 & 1790.14 & 4412.02 & 175.91 \\
77.7910.77 &  38.58 &  983.43 & 5202.34 & 114.76 \\
77.7912.111 &  47.41 &  604.76 & 4155.77 & 135.12 \\
77.7914.74 &  70.70 & 1079.15 & 3933.14 & 227.86 \\
 \hline
 \end{tabular}
 \end{table}

\subsection{The mid-infrared colour of sequence E stars}

It has recently been shown \citep{dust} that variable red giants belonging to sequence D have a mid-infrared excess when compared to similar red giants without the Long Secondary Period that characterises sequence D stars. This mid-infrared excess is assumed to arise from circumstellar dust associated the presence of the LSP.  Since the sequence E stars are close binary systems, some of which have quite eccentric orbits \nocite{ogleellipsoidal} (Soszy\'nski et al. 2004b), there is a possibility that these systems could have substantial circumstellar disks of dust and gas.

In order to investigate this possibility, we searched for a mid-infrared excess in sequence E stars in the LMC using mid-infrared data from the Spitzer Space Telescope SAGE survey \nocite{blum06mn,sagemn} (Blum et al. 2006; Meixner et al. 2006). The sequence E stars were obtained from the catalogue given by \cite{fraser08}.  None of these objects were detected at 24\,$\mu$m in the SAGE survey but 262 were detected at 8\,$\mu$m. The light curves of these 262 stars were examined and it was found that only 184 could be considered as definite sequence E stars, i.e. ellipsoidal or eclipsing binary systems.  The remainder were variables whose light curve characteristics showed that they belonged to sequence D or the pulsation sequences 1 to 4 \citep[using the notation in][]{fraser08} or they were stars of indeterminate light curve type.

In order to see if the sequence E stars show a mid-infrared excess, we obtained a comparator sample of stars that were similar to the sequence E stars, but without binary companions.  Such stars are the normal, non-varying field red giants in the LMC.  Since the sequence E stars that were detected at 8\,$\mu$m have $12.5 <\ K\ < 14$ and $0.6 <\ J-K\ < 1.4$, we selected comparison sources from the SAGE survey with these characteristics along with the requirement of an 8\,$\mu$m detection.  An examination of this sample of $\sim$40000 objects showed that sources fainter than $K\ = 13.5$ were near the 8\,$\mu$m faint detection limit and that this significantly biased the population of detected stars in favour of objects with a higher $K$-[8] colour.  We therefore omitted such stars and we compared sequence E stars and non-varying field red giants only in the interval $12.5 <\ K\ < 13.5$.  Finally, all variable stars in the catalogue of \citet{fraser08} were removed from the list of comparison sources.  This left 144 sequence E stars and $\sim$33000 non-varying field red giants.  The $K$-[8] colour distributions of these two samples were then examined to search for an 8\,$\mu$m mid-infrared excess.  A two sample K--S test gives a probability of up to 0.89 that the sequence E stars and non-varying field red giants come from the same underlying distribution in $K$-[8] colour. In other words, there is no evidence that sequence E variability generally leads to a significant amount of circumstellar dust and a mid-infrared excess.

\section{Discussion and Conclusions}

With new velocity curves, we now have strong evidence that the sequence E variation is indeed caused by ellipsoidal variability. This is shown by the doubling of the light curve with respect to the velocity curve in these stars, as in Figs.~\ref{phased1} and~\ref{phased2}.

However, as we showed in \cite{seqDpaper}, the phased light curves of sequence D variables -- the stars with Long Secondary Periods -- do not show this doubling phenomenon (see fig.~3 of that paper). Although it has been suggested that the sequence D and E variations may have a common origin due to their proximity in the period--luminosity plane \nocite{ogleellipsoidal} (Soszy\'nski et al. 2004b), and the possible existence of ellipsoidal shapes in their residual light curves \citep{soszynski07}, no sequence D star has so far been found whose light executes two cycles during one velocity cycle. This seems to rule out ellipsoidal variation in sequence D stars. 

A further difference between the stars on sequences D and E is demonstrated by their respective velocity amplitudes. The sequence E stars of the current sample show full velocity amplitudes of at least $15\ \rm{km\,s^{-1}}$ -- some much larger -- as can be seen in Fig.~\ref{vhist}. However as we showed in \cite{seqDpaper}, stars with LSPs have much smaller velocity amplitudes, typically around $3.5\ \rm{km\,s^{-1}}$. This marked difference in the size of the velocity variations between sequence D and E -- which can be seen in Fig.~\ref{vamp-p} -- supports our assertion that they are caused by different mechanisms.

The sequence E stars, known binaries, show an expected spread in companion mass estimates. However this distribution seems very different for the sequence D stars. If treated as binaries, sequence D stars have companions that are small and absurdly similar ($\sim 0.09\ \rm{M_{\odot}}$). Additionally, the angles of periastron of sequence D stars are heavily biased towards large values, whereas we would expect a uniform distribution for binaries. Unfortunately we cannot compare the distribution of angle of periastron for the current sequence E sample: as most of these stars have almost circular orbits, angle of periastron is poorly defined (see the large errors in Table~\ref{orbital}).

The distribution of eccentricity is very different for sequence E and D stars, when the latter are treated as binaries. Most of the red giant binaries on sequence E have low-eccentricity orbits, as expected for their evolutionary state. However \cite{seqDpaper} showed that the sequence D stars have a significantly higher eccentricity, if we assume they are binaries (the mean eccentricity of that sample is 0.3). A two-sample K--S test gives a probability of less than $2.1 \times 10^{-6}$ that the sequence D and E eccentricity distributions come from the same underlying distribution. 

Further evidence that sequence D stars are not ellipsoidal variables is given by the relation between luminosity and velocity amplitude in those stars. In Fig.~\ref{k-vamp}, we showed that the sequence E stars generally populate an area delineated by a maximum velocity amplitude--luminosity relation, as expected for ellipsoidal variables. However, \cite{seqDpaper} showed that the sequence D stars do not show any correlation between these properties.

Finally, we have searched for a mid-infrared excess in sequence E stars since such an excess was found among the sequence D stars. No mid-infrared excess was found, demonstrating another difference from the sequence D stars.

In summary, we have demonstrated several significant differences between stars on sequences D and E. Most particularly, we have shown that Long Secondary Periods -- the sequence D variation -- are not caused by ellipsoidal variability, unlike the variation of the sequence E stars. The cause of Long Secondary Periods remains, at this time, a mystery. 

\section*{Acknowledgments}

We are grateful for the multiple allocations of VLT service time over several years for this extended series of observations (program identifiers 072.D-0387, 074.D-0098, 075.D-0090 and 076.D-0162).

This paper utilizes public domain data obtained by the MACHO Project, jointly funded by the US Department of Energy through the University of California, Lawrence Livermore National Laboratory under contract No. W-7405-Eng-48, by the National Science Foundation through the Center for Particle Astrophysics of the University of California under cooperative agreement AST-8809616, and by the Mount Stromlo and Siding Spring Observatory, part of the Australian National University.

This publication makes use of data products from the Two Micron All Sky Survey, which is a joint project of the University of Massachusetts and the Infrared Processing and Analysis Center/California Institute of Technology, funded by the National Aeronautics and Space Administration and the National Science Foundation.

\bibliographystyle{mn2e}
\bibliography{bibliographynew}

\label{lastpage}

\end{document}